\documentclass[showkeys,amsmath,amssymb,twocolumn,floatfix]{revtex4}
\usepackage[dvips]{graphics,graphicx}
\input{epsf}

\usepackage{txfonts}

\begin{document}

\title{Entanglement and localization of wavefunctions}
\author{O.~Giraud$^{1,2}$, J.~Martin$^{1,2,3}$ and B.~Georgeot$^{1,2}$}
\affiliation{\mbox{$^{1}$Universit\'e de Toulouse, UPS, Laboratoire
de Physique Th\'eorique (IRSAMC), F-31062 Toulouse, France} \\
\mbox{$^{2}$CNRS, LPT (IRSAMC), F-31062 Toulouse, France} \\
\mbox{$^{3}$Institut de Physique Nucl\'eaire, Atomique et de
Spectroscopie, Universit\'e de Li\`ege, 4000 Li\`ege, Belgium}}

\date{July 20, 2009}

\begin{abstract}
We review recent works that relate entanglement of random vectors to
their localization properties. In particular, the linear entropy is
related by a simple expression to the inverse participation ratio,
while next orders of the entropy of entanglement contain information
about e.g.\ the multifractal exponents.  Numerical simulations show
that these results can account for the entanglement present in
wavefunctions of physical systems.
\end{abstract}


\maketitle

\section{Introduction}

Quantum mechanics has always seemed puzzling since its first
construction in the first half of the XXth century.  Many properties
are different from the world of classical physics in which our
intuition is built.  The development of quantum information science
in the last decades has exemplified this aspect.  Indeed, it was
realized that it is in principle possible to exploit the features of
quantum mechanics to treat information in a different way from what a classical
computer would do. In this context, the
specific properties of quantum mechanics are put forward as new
resources which enable to treat information in completely new ways.

One of the most peculiar properties of quantum mechanics is
entanglement, that is the possibility to construct quantum states of
several subsystems that cannot be factorized into a product of
individual states of each subsystem. Such entangled states are the
most common in quantum mechanics, and they display correlations
which cannot be seen in a classical world, exemplified by e.g.\ the
Einstein-Podolsky-Rosen ``paradox''.  Entanglement is also a
resource for quantum information (see \cite{nielsen} and references
therein),
 and has been widely studied as such in the past few years.

Despite intensive work, entanglement remains a somewhat mysterious
property of physical systems.  The structure of entanglement of
systems even with small numbers of particles is hard to
characterize. Even properly measuring the entanglement present in a
system is difficult for mixed states. This is all the more important
since recent results have shown that (at least for pure states) if a
process creates a sufficiently low level of entanglement, it can be
simulated efficiently by a classical computer \cite{jozsa}.  This
gives a limit on the speedup over classical computation a quantum
computer can achieve, and also gives rise to interesting proposals
for building classical algorithms simulating weakly entangled
quantum systems \cite{cirac}.

In this paper, we review recent results we obtained (details can be
found in \cite{GMG1,GMG2}), which concern the relationship of
entanglement to localization properties of a quantum state.  Our
strategy is to consider $n$-qubit systems, and to study entanglement
of quantum states relative to their localization properties in the
$2^n$-dimensional Hilbert space in the computational basis. We
obtain analytical results for {\em random states}, that is ensemble
of quantum states sharing some properties.  Such random states have
been recently studied in the literature. They are interesting in
themselves, since it has been shown for example in quantum
information that they are useful in various quantum protocols
\cite{random}.  This motivated a recent activity in the quantum
information community to try and produce efficiently such random
vectors or random operators through quantum algorithms
\cite{emerson}, and to characterize their entanglement properties
\cite{ranvec}. In addition to their intrinsic usefulness, random
states are important since they can describe typical states of a
"complex" system.  For example, it has been known for some time now
that random vectors built from Random Matrix Theory (RMT) can
describe faithfully the properties of quantum Hamiltonian systems
whose classical limit is chaotic, and more generally of many complex
quantum systems \cite{houches}.  Such random vectors are ergodic,
and the entanglement they contain has been calculated some time ago
\cite{Lub, Pag}.  However, in many quantum systems, the
wavefunctions are not ergodic but localized.  This can correspond to
electrons in a disordered potential, which are exponentially
localized due to Anderson localization.  It can also be seen in
many-body interacting systems, where the presence of a moderate
interaction can lead to states partially localized in energy.
 Some systems are in a well-defined sense neither ergodic
neither localized: they correspond to e.g.\ states at the Anderson
transition between localized and delocalized states, and can show
multifractal properties \cite{mirlin}.

In this paper, we calculate the amount of entanglement
present in ensembles of random vectors displaying these various degrees
of localization.  Besides generalizing the result for RMT-type random
vectors, this gives the entanglement present in a ``typical state'' of
such localized or partially localized systems.
This enables to estimate the complexity of
simulating such systems on classical computers, but also sheds light
on the entanglement itself, since in these cases it is related through
simple formulas to quantities characterizing the degree of
localization of the system.

Our results show that for random vectors which are localized on the
computational basis, the linear entropy which approximates the
amount of entanglement in the vector is simply related to the
Inverse Participation Ratio (IPR), a popular measure of
localization.  The next term in the approximation is related to
higher moments, and in particular to the multifractal exponents for
multifractal systems.  In order to assess the usefulness of these
results to physical systems, we compare them to the entanglement
numerically computed for several models. After a general discussion
on entanglement of random vectors (section II), we consider the
entanglement of one qubit with the others (section III), and give
explicitly the first and second order of the expansion of the
entropy of entanglement around its maximum. Section IV generalizes
these results to other bipartitions, and section V compares the
formula obtained with the numerical results for two physical
systems. Section VI considers the physically important case of
vectors localized not on a random subset of the basis vectors, but
on a subset composed of adjacent basis vectors (that is the states
are localized on computational basis states which are adjacent when
the basis vectors are ordered according to the number which labels
them), showing that the results become profoundly different. Section
VII presents the conclusions.

\section{Entanglement of random vectors}

Random vectors are ensembles of vectors whose components are
distributed according to some probability distribution. If for
example the system considered is composed of $n$ qubits, the Hilbert
space is of dimension $N=2^n$.
If the two states of a qubit are denoted $|0\rangle$
and $|1\rangle$, each state in the computational basis corresponds
to a sequence of $0$ and $1$ and thus can be labelled naturally by a
number between $0$ and $2^n-1$, and quantum states can be expanded
as $|\psi\rangle=\sum_i \psi_i |i\rangle$.
Random vectors distributed
according to the uniform measure on the $N$-dimensional sphere
describe typical quantum states of the $n$ qubits.  Such states are
ergodically distributed in the computational basis,
and their entanglement has already been
studied in \cite{Lub,Pag}. In this paper, we are interested
in random vectors which are not ergodically distributed.
Ensembles of such states will be characterized by localization
properties. The simplest example of such localized random vectors
can be constructed by taking $M$ components ($M<N$) with equal
amplitudes and uniformly distributed random phases, and setting all
the others to zero.  The random vectors will all be exactly
localized on $M$ basis states.
%(e^{i\theta_1}/\sqrt{M},...,e^{i\theta_M}/\sqrt{M})$.
A more physically relevant example consists in still choosing $M<N$
nonzero components, and giving them the distribution of column
vectors of $M\times M$ random unitary matrices drawn from the
Circular Unitary Ensemble of random matrices (CUE vectors).  In
general our result will be averaged both over the distribution of
the nonzero components and the position of these nonzero components
in the computational basis. This corresponds to classes of random
vectors sharing the same localization length.  Our results will in
fact generalize to any such distribution of random vectors whose
localization properties are fixed.  In addition, we shall see that
if we impose that the distribution of the indices $i$ of nonzero
components $\psi_i$ is such that they are always adjacent in the
computational basis (i.e.\ the indices $i$ are consecutive
integers), the results change drastically.

The localization  properties of the random vectors can be probed using
the moments of the distribution
\begin{equation}
p_q=\sum_{i=1}^N|\psi_i|^{2q}.
\end{equation}

The second moment is $p_2=1/\xi$, where
$\xi$ is the Inverse Participation Ratio
(IPR) which is often used in the mesoscopic physics literature to measure
the localization length. Indeed, for a state uniformly spread on
exactly $M$ basis vectors, one has $\xi=M$.
  The scaling of $p_2$ and higher moments with
the size also probes the multifractal properties of the wavefunction.

The random states we consider are built on the $N$-dimensional
Hilbert space of a $n$-qubit system with $N=2^n$. We are interested
in bipartite entanglement between subsystems defined by different
partitions of the $n$ qubits into two sets. In general, bipartite
entanglement of a pure state belonging to a Hilbert space ${\mathcal
H}_A\otimes{\mathcal H}_B$ is measured through the entropy of
entanglement, which has been shown to be a unique entanglement
measure \cite{PopRoh}. We consider pure states belonging to
${\mathcal H}_A\otimes{\mathcal H}_B$ where ${\mathcal H}_A$ is a
set of $\nu$ qubits and ${\mathcal H}_B$ a set of $n-\nu$ qubits. If
$\rho_A = \textrm{tr}_B |\psi\rangle\langle\psi|$
is the density matrix obtained by tracing out subsystem
$B$, then the entropy of entanglement of the state $\psi$ with
respect to the bipartition $(A,B)$ is the von Neumann entropy of
$\rho_A$, that is $S=-\textrm{tr}(\rho_A\log_2\rho_A)$.

\section{Entanglement of one qubit with all the others}

To obtain an approximation for the entropy, one can expand $S$
around its maximal value.  In the case of the partition of the $n$
qubits into $1$ and $n-1$ qubits, the entropy can be written as a
function of $\tau$, with
\begin{equation}
\tau =4\det\rho_A
\end{equation}
(in the case of 2 qubits this quantity is called the {\it tangle}
and corresponds to the square of the generalized concurrence
\cite{RunCav}). One has
\begin{equation}\label{S}
S(\tau)=h\left(\frac{1+\sqrt{1-\tau}}{2}\right),
\end{equation}
where $h(x)=-x\log_2 x-(1-x)\log_2(1-x)$. The series expansion of
$S(\tau)$ up to order $m$ in $(1-\tau)$ reads
\begin{equation}
\label{expansionS} S_m(\tau)=1-\frac{1}{\ln
2}\sum_{n=1}^{m}\frac{(1-\tau)^n}{2n(2n-1)}.
\end{equation}

The first order corresponds to $\tau$ itself up to constants
and its average over
the choice of the $(1,n-1)$ partition is known as the linear entropy
or Meyer-Wallach entanglement $Q$ \cite{MW}. Our results show that for
our class of random vectors, the average linear entropy is given by
\begin{equation}
\label{tau1} \langle\tau\rangle=\frac{N-2}{N-1}(1-\langle
p_2\rangle)=\frac{N-2}{N-1}(1-\langle
1/\xi\rangle).
\end{equation}

This formula was obtained first by considering a random vector which
is nonzero only on $M$ basis vectors among $N$, and summing
explicitly the combinatorial terms.  It can also be obtained in a
more general setting by taking $M=N$ and summing up all the
localization properties of the vector in the IPR $\xi$ alone. For
any $(1,n-1)$ partition of the $n$ qubits, the components of the
vector can be divided in two sets according to the value of the
first qubit. Assuming no correlation among these sets enables to get
Eq.~(\ref{tau1}) (details on the calculations can be found in
\cite{GMG1}).

It is interesting to compare this formula with a similar one obtained
in \cite{viola}
using different assumptions,
in particular without average over random phases. The formula obtained relates
entanglement to the mean inverse participation ratio
calculated in three different bases,
a quantity that is more general but often delicate to evaluate.
In our case, the additional assumption
of random phases enables to obtain a formula which involves only the
IPR in one basis,
a quantity that can be easily evaluated in many cases.  For example,
it enables to compute readily the entanglement for localized CUE vectors.
However there are instances of systems (e.g.\ spin systems)
where these different formulas give the same results.

In particular, our formula (\ref{tau1}) allows to compute
$\langle \tau\rangle$ e.\ g.\
for a CUE vector localized on $M$ basis vectors;
in this case  $\xi=(M+1)/2$, and
we get
\begin{equation}
\label{QnadjCUE}
\langle \tau\rangle=\frac{M-1}{M+1}\frac{N-2}{N-1}.
\end{equation}
In \cite{Lub},  $\langle \tau\rangle$ was calculated for
non-localized CUE vectors of length $N$, giving
$\langle \tau\rangle=(N-2)/(N+1)$. Consistently, our formula yields
the same result if we take $M=N$. For a
vector with constant amplitudes and random phases on $M$ basis vectors,
$\xi=M$ and
\begin{equation}
\langle \tau\rangle=\frac{M-1}{M}\frac{N-2}{N-1}.
\end{equation}
Thus the first order of the expansion, which gives
the main features of the entanglement, has very simple
expressions in term of system parameters.

The next order in the expansion (\ref{expansionS}) can be obtained
by similar methods that we do not detail here (see \cite{GMG2} for
details); summing up all terms involved in $\tau^2$ we get
\begin{eqnarray}
\label{tau2}
\langle\tau^2\rangle&=&N(N-2)(N^2-6N+16)c_{1111}\\
&& + 4N(N-2)(N-4)c_{211}+4N(N-2)c_{22}\nonumber
\end{eqnarray}
with
\begin{eqnarray}\label{cs}
c_{22}&=&\frac{\langle p_2^2\rangle-\langle p_4\rangle}{N(N-1)},\
c_{211}=\frac{\langle p_2\rangle-\langle p_2^2\rangle-2\langle p_3\rangle+2\langle p_4\rangle}{N(N-1)(N-2)},\nonumber\\
c_{1111}&=&\frac{1-6\langle p_2\rangle+8\langle p_3\rangle+3\langle
p_2^2\rangle-6\langle p_4\rangle}{N(N-1)(N-2)(N-3)}.
\end{eqnarray}

This gives the next order of the entropy of entanglement in terms of
the moments up to order 4 of the vector.  What this means is that at
this order, the average entanglement of random vectors with fixed
moments will be related to them through (\ref{tau2}).  Although more
complicated than (\ref{tau1}), the formula indicates that e.g.\ for
states having multifractal properties, since moments scale with
system size according to quantities called multifractal exponents,
the behavior of the entanglement at this order will be also
controlled by these multifractal exponents.

The $n$th order of the expansion (\ref{expansionS}) can similarly be
obtained and has been derived in \cite{GMG2}. It is interesting to
note that in the case of a CUE random vector of size $N$,
resummation of the whole series for $S(\tau)$ yields, after some
algebra,
\begin{equation}
\langle S(\tau)\rangle=\frac{1}{\ln
2}\sum_{k=N/2+1}^{N-1}\frac{1}{k},
\end{equation}
which has been obtained earlier by a different method \cite{Pag}.

A general conclusion obtained from these formulas is that the
entanglement associated to such bipartition $(1,n-1)$ goes to the
maximal value for large $N$ and large $\xi$, even if $\xi$ grows
more slowly than $N$. For fixed $\xi$, it tends for large $N$ to a
constant nonzero value which depends on $\xi$. We will see in
Section VI that this result can change drastically if we impose a
localization on fixed locations in Hilbert space.

\section{Entanglement of random vectors: other partitions}

Up to now we have considered the entanglement of one qubit with all
the others, i.e.\ the $(1,n-1)$ partition of $n$ qubits. What about
bipartite entanglement relative to other bipartitions $(\nu,n-\nu)$,
where $\nu$ is any number between $1$ and $n-1$? In this case, it is
convenient to define the linear entropy as
$S_L=\frac{d}{d-1}(1-\textrm{tr}\rho_A^2)$, where $d=\dim{\mathcal
H}_A\leq \dim{\mathcal H}_B$. The scaling factor is such that $S_L$
varies in $[0,1]$.

A similar calculation as above enables then to obtain the first order
of the mean von Neumann entropy.  It is given by
\begin{equation}
\label{taugeneral} \langle S\rangle \approx \nu-\frac{2^{\nu}-1}{2\ln 2}
\left(1-\frac{N-2^{\nu}}{N-1}\Big\langle\frac{1}{\xi}\Big\rangle\right),
\end{equation}
with $p_2=1/\xi$, which generalizes Eq.~(\ref{tau1}).

Higher-order terms can be obtained as well, although the
calculations become tedious. To this end, the entropy
$S=-\textrm{tr}(\rho_A\log_2\rho_A)$ is expanded around the maximally mixed
state $\rho_0=\mathbf{1}/2^{\nu}$, as
\begin{equation}
\label{Snu} S=\nu+\frac{1}{\ln 2}\sum_{n=1}^{\infty}
\frac{(-2^\nu)^{n}}{n(n+1)}\textrm{tr}((\rho_A-\rho_0)^{n+1}),
\end{equation}
and the traces can be evaluated as sums over correlators of higher moments
\cite{GMG2}.

We remark that again the linear entropy (\ref{taugeneral}) tends to
the maximal possible value when $N$ and $\xi$ become large, as for
the $(1,n-1)$ partition.

\section{Entanglement of random vectors: application to physical systems}

In order to test these results on physical systems, we compared
them to numerical results obtained from different models.

\begin{figure}
\centerline{\includegraphics[width=3in]{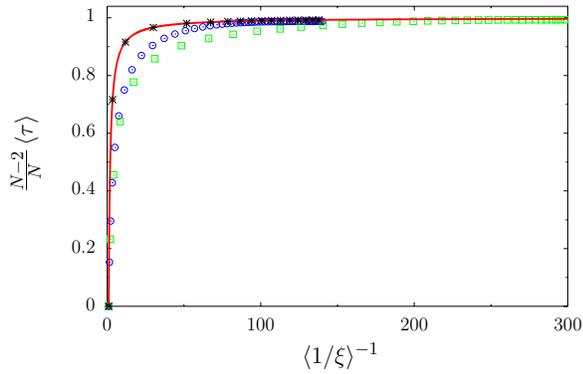}} \caption[]{Scaled
mean linear entropy $\langle \tau \rangle (N-2)/N$ of (\ref{hamil})
vs mean IPR for $\delta=\Delta_0$, $n=10$ (blue circles) and $n=11$
(green squares). Red line is the theory, stars the data for $n=10$
with random shuffling of components (from \cite{GMG1}).} \label{xi}
\end{figure}

The first one corresponds to a diagonal Hamiltonian matrix
to which a two-body interaction is added.

\begin{equation}
H = \sum_{i} \Gamma_i \sigma_{i}^z + \sum_{i<j} J_{ij}
\sigma_{i}^x \sigma_{j}^x
\label{hamil}
\end{equation}

This system can describe a quantum computer in presence of static
disorder \cite{qchaos}. Here the $\sigma_{i}$ are the Pauli matrices
for the qubit $i$, the energy spacing between the two states of
qubit $i$ is given by $\Gamma_i$, which are randomly and uniformly
distributed in the interval $[\Delta_0 -\delta /2, \Delta_0 + \delta
/2 ]$, and $J_{ij}$ uniformly distributed in the interval $[-J,J]$
represent a random static interaction.  Entanglement of eigenvectors
of this Hamiltonian was already considered in a different context in
\cite{italians}. It is known \cite{qchaos} that in this  model a
transition to quantum chaos takes place for sufficiently large
coupling strength $J$. In this regime, eigenvectors of (\ref{hamil})
are spread over all noninteracting eigenstates (those of
(\ref{hamil}) for $J=0$, which coincide to the computational basis),
but in a certain window of energy, and are distributed according to
the Breit-Wigner (Lorentzian) distribution.  Thus these
wavefunctions are distributed among a certain subset of the
computational basis, although they are not strictly zero outside it,
and the distribution is not uniform, but rather Lorentzian.
Nevertheless, our data show (see Figs.~1,2) that the behavior of the
bipartite entanglement of eigenvectors of this model is well
described by the results (\ref{tau1}) and (\ref{taugeneral}) derived
for random vectors. The agreement becomes very accurate if the
eigenvector components are randomly shuffled to lower correlations.

\begin{figure}
\centerline{\includegraphics[width=3in]{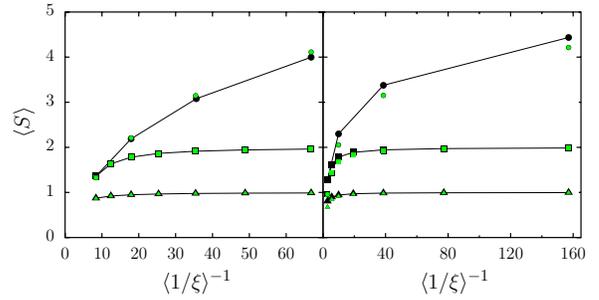}}
\caption[]{Mean entropy of entanglement $S$ for
different bipartitions $(\nu,n-\nu)$ as a function of the mean IPR.
Left: eigenvectors of (\ref{ISRM}) with $\gamma=1/3$; the average is
taken over $10^6$ eigenvectors. Right: eigenvectors of (\ref{hamil})
with $\delta =\Delta_0$ and
$J/\delta=1.5$; average over  $\approx 3\times 10^5$ vectors.
Triangles correspond
to $\nu=1$, squares to $\nu=2$ and circles to $\nu=n/2$, with
$n=4-10$. Black symbols are the theoretical predictions for the mean
value of $S$ (obtained from Eq.~(\ref{taugeneral}) and
green (gray) symbols are the computed mean values of the von Neumann
entropy (from \cite{GMG2}). } \label{firstorder}
\end{figure}

We also considered another model, based on $N\times N$ matrices of the form
\begin{equation}\label{ISRM}
    U_{kl}=\frac{e^{i\phi_k} }{N}\frac{1-e^{2i\pi N \gamma}}{1-e^{2i\pi (k-l+N\gamma)/N}},
\end{equation}
where $\phi_k$ are random variables independent and uniformly
distributed in $[0,2\pi[$ and $\gamma$ is a fixed parameter. This
model introduced in \cite{bogomolny} is the randomized version of a
simple quantum map introduced in \cite{giraud}. The eigenvectors of
(\ref{ISRM}) have multifractal properties \cite{MGG} for rational
$\gamma$. The results of Figs.~2,3 show that again the results for
random vectors describes very well the entanglement for this system
for randomly shuffled components, and that even the first order is
already a good approximation.

\begin{figure}
\centerline{\includegraphics[width=3in]{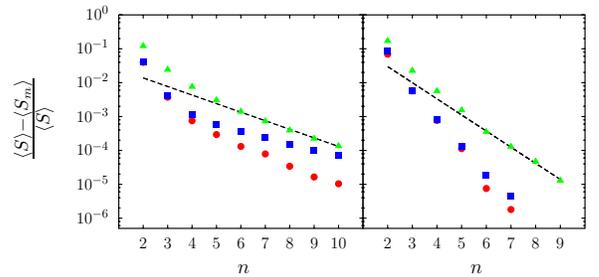}}
\caption[]{Relative difference of the entropy of
entanglement (\ref{S}) and its successive approximations $S_m$
($m=1,2$) with respect to the number of qubits for eigenvectors of
(\ref{ISRM}) for (left) $\gamma=1/3$ and (right) $\gamma=1/7$. The average
is taken over $10^7$ eigenvectors, yielding an accuracy $\lesssim
10^{-6}$ on the computed mean values. Green triangles correspond to
the first order expansion $S_1$, blue squares and red circles to the
second order expansion $S_2$. The difference between the latter two
is that for blue squares $\langle p_2^2 \rangle$ appearing in
Eq.~(\ref{cs}) has been replaced by $\langle p_2 \rangle^2$ yielding
a less accurate approximation. Dashed line is a linear fit yielding
$1-\langle S_1\rangle/\langle S\rangle$ $\sim N^{-0.84}$ for
$\gamma=1/3$ and $N^{-1.58}$ for $\gamma=1/7$ (from \cite{GMG2}). } \label{secondorder}
\end{figure}

\section{Entanglement of adjacent random vectors}

In the preceding sections we discussed formulas for entanglement of
ensembles of random vectors where the components over each basis
vector are independent.  If we relax this assumption, the result may
change.  An important particular case corresponds e.g.\ to random
vectors localized on $M$ computational basis states which are
adjacent when the basis vectors are ordered according to the number
which labels them (again, if the two states of a qubit are denoted
$|0\rangle$ and $|1\rangle$, each state in the computational basis
corresponds to a sequence of $0$ and $1$ and thus can be labelled
naturally by a number between $0$ and $2^n-1$). In this case, we had
to use combinatorial methods; summing all contributions together we
get for the linear entropy of $(1,n-1)$ partitions
\begin{eqnarray}
\label{Qmoy} \langle
\tau \rangle&=&\left[\left(\frac{M-2}{M-1}r_0+\frac{2(2^{r_0}-1)}{M(M-1)}
+\frac{4}{3}\frac{(M+1)(2^n-2^{r_0})}{2^{n+r_0}}\right.\right.\nonumber\\
&&-\left.\left.\frac{1}{M(M-1)}\sum_{r=0}^{r_0-1}\chi_r(m_r)\right)
\left(1-\langle\frac{1}{\xi}\rangle\right) \right]\frac{1}{n},
\end{eqnarray}
where $r_0$ is such that $2^{r_0-1}< M\leq2^{r_0}$ and
$\chi_r(x)=\chi_r(2^{r+1}-x)=x^2-\frac{2}{3}x(x^2-1)/2^r$ for $0\leq
x\leq 2^r$. Equation (\ref{Qmoy}) is an exact formula for $M\leq
N/2$. For fixed $M$ and $n\rightarrow \infty$, $n\langle Q\rangle$
converges to a constant $C$ which is a function of $M$ and $\xi$.
For $M=2^{r_0}$, $r_0<n$, Eq.~(\ref{Qmoy}) simplifies to
\begin{eqnarray}
\label{Qmoyapprox} \langle
\tau\rangle&=&\left[\left(\frac{(r_0+\frac{4}{3})M^2-2(r_0-1)M-\frac{10}{3}}{M(M-1)}\right.\right.\nonumber\\
&&-\left.\left.\frac{4(M+1)}{3
N}\right)\left(1-\langle\frac{1}{\xi}\rangle\right)\right]
\frac{1}{n}.
\end{eqnarray}
Numerically, this expression with $r_0=\log_2M$ gives a very good
approximation to Eq.~(\ref{Qmoy}) for all $M$.

\begin{figure}
\centerline{\includegraphics[width=3in]{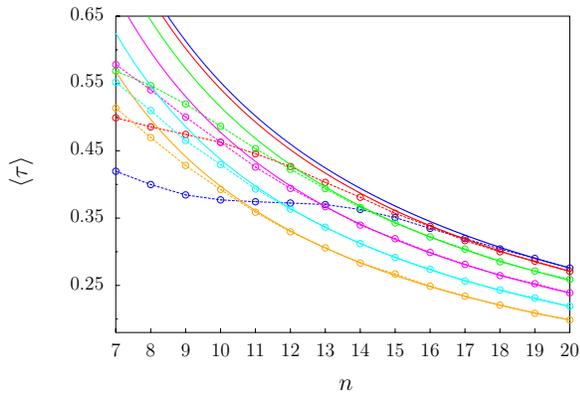}} \caption[]{Mean
linear entropy $\langle \tau \rangle$ of partitions $(1,n-1)$ vs
number of qubits for the one-dimensional Anderson model with
disorder from top to bottom $w=0.2$ (blue), $0.5$ (red), $1.0$
(green), $1.5$ (magenta), $2.0$ (cyan), and $2.5$ (orange). Average
is over 10000 eigenstates. Solid lines are the $C/n$ fits of the
tails (from \cite{GMG1}).
%(in practice we took a CUE vector of length $20\xi+1$ so that the exponential is truncated at $\pm 10\xi$).
}
\label{Qanderson}
\end{figure}

Equation (\ref{Qmoy}) is exact for e.g.\ uniform and CUE vectors,
and can be applied even if the vector is not strictly zero outside a
$M$-dimensional subspace.  Indeed, for $N$-dimensional CUE vectors
with exponential envelope $\exp(-x/l)$, $\langle Q\rangle$ is in
excellent agreement with Eq.~(\ref{Qmoy}) with $\xi=l$ and $M=2\xi$.

In order to compare these findings to those of a physical system
with such a property of localization on adjacent basis vectors,
Fig.~4 shows the theory Eq.~(\ref{Qmoy}) together with the entropy
for the one-dimensional Anderson model.  This model corresponds to a
one-dimensional chain of vertices with nearest-neighbor coupling and
randomly distributed on-site disorder, described by the Hamiltonian
$H_0+V$. Here $H_0$ is a diagonal operator whose elements
$\epsilon_i$ are Gaussian random variables with variance $w^2$, and
$V$ is a tridiagonal matrix with non-zero elements only on the first
diagonals, equal to the coupling strength, set to $1$.  It is known
that eigenstates of this system, which modelizes electrons in a
disordered potential, have envelopes of the form $\exp(-|x-x_0|/l)$,
where $l$ is the localization length.  It was shown in
\cite{pomeransky} that this model can be simulated efficiently on a
quantum computer, and the wavefunction of the computer during the
algorithm will be localized on adjacent basis vectors, which
correspond to the position of vertices. Figure 4 shows that the
asymptotic behavior of the linear entropy of the eigenstates (with
all correlations left between components, i.e.\ no random shuffling)
is well captured by Eq.~(\ref{Qmoy}).

Thus random vectors localized on adjacent basis vectors correspond
to a drastically different behavior compared to the vectors of
section III~: indeed, for fixed $\xi$ the entanglement (at least the
linear entropy) always tends to zero for large $N$, even if it does
it rather slowly (as $\sim 1/\ln N$).

\section{Conclusion}

The results above indicate that the entanglement properties
 of random vectors can
be directly related to the fact that they
are localized, multifractal
or extended. The numerical simulations for different physical systems
show that these results obtained for random vectors describe
qualitatively the entanglement present in several physical systems,
and reproduce it accurately if correlations between components
of the vector are averaged out.

Thus the results are interesting to predict the amount of
entanglement present in random vectors, and also can be applied to
physical systems for which such random vectors describe typical
states. This gives insight on the difficulty to simulate classically
such systems, since systems with low amounts of entanglement can be
simulated classically efficiently. This also can be applied to
estimate the changes in entanglement at a quantum phase transition
\cite{spins}, in particular for the Anderson transition between
localized and extended states (see \cite{GMG2} for more details).
Additionally, this gives also insight on the nature of entanglement
itself by relating it to simple physical properties of the system.

\acknowledgements We thank CalMiP for access to their
supercomputers. This work was supported by the Agence Nationale de
la Recherche (project ANR-05-JCJC-0072 INFOSYSQQ), the Institut de Physique
of CNRS (project PEPS-PTI) and the European program EC IST FP6-015708 
EuroSQIP. J.M.\ thanks the Belgian F.R.S.-FNRS for financial support.


\begin{thebibliography}{99}
\bibitem{nielsen}M.~A.~Nielsen and I.~L.~Chuang, {\em Quantum computation
                  and quantum information}, Cambridge Univ. Press, 2000.
\bibitem{jozsa} R. Jozsa and N. Linden, Proc. R. Soc. London Ser. A
{\bf 459}, 2011 (2003); G.~Vidal,
Phys. Rev. Lett. {\bf 91}, 147902 (2003).
\bibitem{cirac}
G.~Vidal, Phys. Rev. Lett. {\bf 91}, 147902 (2003); F.~Verstraete,
D.~Porras, and J.~I.~Cirac, Phys. Rev. Lett. {\bf 93}, 227205
(2004).
\bibitem{GMG1} O. Giraud, J. Martin, and B. Georgeot, Phys. Rev. A  {\bf 76},
042333 (2007).
\bibitem{GMG2} O. Giraud, J. Martin, and B. Georgeot, Phys. Rev. A  {\bf 79},
032308 (2009).
\bibitem{random} A.~Harrow, P.~Hayden and D.~Leung,
Phys. Rev. Lett. {\bf 92}, 187901 (2004);
P.~Hayden, D.~Leung, P.~Shor and A.~Winter, Commun. Math. Phys. {\bf 250},
371 (2004); C.~H.~Bennett, P.~Hayden,
D.~Leung, P.~Shor and A.~Winter, IEEE Trans. Inf. Theory {\bf 51}, 56 (2005).
P.~Cappellaro, J.~Emerson, N.~Boulant, C.~Ramanathan and
D.~G.~Cory, Phys. Rev. Lett. {\bf 94}, 020502 (2005).
\bibitem{emerson} J.~Emerson, Y.~S.~Weinstein, M.~Saraceno, S.~Lloyd and
D.~S.~Cory, Science {\bf 302}, 2098 (2003);
Y.~S.~Weinstein and C.~S.~Hellberg, Phys. Rev. Lett. {\bf 95},
030501 (2005).
\bibitem{ranvec} A.J.~Scott, Phys. Rev. A {\bf 69}, 052330 (2004);
H.-J.~Sommers and K.~Zyczkowski, J. Phys. A {\bf 37}, 8457 (2004);
O.~Giraud, J. Phys. A {\bf 40}, 2793 (2007); O.~Giraud,
J. Phys. A {\bf 40}, F1053 (2007); M.~Znidaric,
J. Phys. A {\bf 40}, F105 (2007); M.~Znidaric, T.~Prosen, G.~Benenti
and G.~Casati, J. Phys. A {\bf 40}, 13787 (2007);
P.~Facchi, U.~Marzolino, G.~Parisi, S.~Pascazio and A.~Scardicchio,
Phys. Rev. Lett. {\bf 101}, 050502 (2008).
\bibitem{houches} {\it Chaos and quantum physics},
Proceedings of the 52th Les Houches Summer School, Eds. M.-J. Giannoni,
A. Voros and J. Zinn-Justin (North-Holland, Amsterdam,1991).
\bibitem{Lub} E.~Lubkin, J. Math. Phys. (N.Y.) {\bf 19},
1028 (1978).
\bibitem{Pag} D.~N.~Page, Phys. Rev. Lett. {\bf 71}, 1291 (1993).
\bibitem{mirlin} A.~D.~Mirlin, Phys. Rep. {\bf 326}, 259 (2000);
F.~Evers and A.~D.~Mirlin, arXiv:0707.4378.
\bibitem{PopRoh} S.~Popescu and D.~Rohrlich, Phys. Rev. A {\bf 56}, R3319
 (1997).
\bibitem{RunCav} P.~Rungta and C.~M.~Caves,
Phys. Rev. A {\bf 67}, 012307  (2003).
%\bibitem{DonHorRud} M.~J.~Donald, M.~Horodecki and O.~Rudolph , J. Math. Phys.
%{\bf 43}, 4252 (2002).
%\bibitem{MunJam01} W.~J.~Munro, D.~F.~V.~James, A.~G.~White, and P.~G.~Kwiat,
%Phys. Rev. A {\bf 64}, 030302 (2001).
\bibitem{MW}A.~D.~Meyer and N.~R.~Wallach, J. Math. Phys. {\bf 43},
 4273 (2002). G.~K.~Brennen, Quant. Inf. Comp. {\bf 3} 619 (2003).
\bibitem{viola}L.~Viola and W.~G.~ Brown, J. Math. Phys. {\bf 43},
8109 (2007); W. G. Brown, L. F. Santos, D. J. Starling and L. Viola,
Phys. Rev. E {\bf 77}, 021106 (2008).
\bibitem{qchaos} B.~Georgeot and D.~L.~Shepelyansky,
Phys. Rev. E {\bf 62}, 3504 (2000); {\it ibid.} {\bf 62}, 6366
(2000).
\bibitem{italians}
C.~Mejia-Monasterio, G.~Benenti, G.~G.~Carlo and G.~Casati,
 Phys. Rev. A {\bf 71}, 062324 (2005).
\bibitem{bogomolny} E.~Bogomolny and C.~Schmit,
Phys. Rev. Lett. {\bf 93}, 254102 (2004).
\bibitem{giraud} O.~Giraud, J.~Marklof and S.~O'Keefe, J. Phys. A
{\bf 37}, L303 (2004).
\bibitem{MGG}
J. Martin, O. Giraud, and B. Georgeot, Phys. Rev. E  {\bf 77},
R035201 (2008).
\bibitem{pomeransky} A.~A.~Pomeransky and D.~L.~Shepelyansky,
Phys. Rev. A {\bf 69}, 014302 (2004); O.~Giraud, B.~Georgeot and D.~L.~Shepelyansky,
Phys. Rev. E {\bf 72}, 036203 (2005).
\bibitem{spins} L. Amico, R. Fazio, A. Osterloh and V. Vedral, Rev. Mod. Phys. {\bf 80}, 517 (2008).

\end{thebibliography}
\end{document}